# Low microwave loss in deposited Si and Ge thin-film dielectrics at single-photon power and low temperatures

Cameron J. Kopas[1], Justin Gonzales[1], Shengke Zhang[1], D. R. Queen[2], Brian Wagner[2], Mcdonald Robinson[3], James Huffman[3], and Nathan Newman[1]

[1]Materials Program, Arizona State University, Tempe, Arizona, 85287, USA
[2]Northrop Grumman Corporation, Mission Systems, Linthicum, Maryland 21090, USA
[3]Lawrence Semiconductor Research Laboratory, Tempe, Arizona, 85282, USA

**Abstract**

Our study shows that deposited Ge and Si dielectric thin-films can exhibit low microwave losses at single-photon powers and sub-Kelvin temperatures (≈40 mK). This low loss enables their use in a wide range of devices, including low-loss coplanar, microstrip, and stripline resonators, as well as layers for device isolation, interwiring dielectrics, and passivation in microwave and Josephson junction circuit fabrication. We use coplanar microwave resonator structures with narrow trace widths of 2-16 μm to maximize the sensitivity of loss tangent measurements to the interface and properties of the deposited dielectrics, rather than to optimize the quality factor. In this configuration, thermally-evaporated ≈1 μm thick amorphous germanium (*a*-Ge) films deposited on Si (100) have a single photon loss tangent of $1\text{-}2\times10^{-6}$ and, 9 μm-thick chemical vapor deposited (CVD) homoepitaxial single-crystal Si has a single photon loss tangent of $0.6\text{–}2\times10^{-5}$. Interface contamination limits the loss in these devices.

Microwave resonators fabricated for use in superconducting quantum computing and sensing applications are almost always fabricated in the co-planar configuration on bulk, high-purity single-crystal Si and sapphire ($Al_2O_3$) dielectrics[1–4]. The resonator performance at low temperatures and single-photon powers is limited by dielectric and two-level system losses in the dielectric material, and at the metal-air, metal-dielectric, and dielectric-air interfaces[5–8].

We suggest several-micron-thick high-purity CVD homoepitaxial silicon as an alternative dielectric for co-planar structures for the following reasons. These deposited films are available in very pure form with uncompensated carrier concentrations as low as $10^{12}$ cm$^{-3}$ and structural defect densities less than 1 cm$^{-2}$. Epitaxial Si layers have proven to have utility in high-power Si device applications over their bulk Si counterparts based on high breakdown voltage. Such layers are believed to have a different defect composition than float-zone Si wafers, which have of order $5\times10^{15}$ cm$^{-3}$ oxygen and carbon contaminants and a significantly higher number of electrically inactive defects (i.e., do not degrade performance at room temperature)[9]. Such bulk Si defects could undergo multi-atom configuration changes at low temperatures and thus contribute to loss and noise in quantum-based computing and sensing applications.

While the favored substrates for these low-temperature applications are single crystal material, other dielectrics, including $SiO_2$, $Si_3N_4$, and $SiO_xN_y$, are used as isolation, interwiring, and passivation layers in microwave device and Josephson junction circuit processes. Unfortunately, these materials have high loss tangents[1] of $0.1–3\times10^{-3}$. Materials for device isolation and passivation are typically deposited using plasma-enhanced chemical vapor deposition (PECVD), which requires temperatures of ≈ 350–400 °C to produce electrically-insulating low-defect density material. However, for superconductor device fabrication, temperature limits[10] (150-175 °C) are imposed to prevent thermally induced changes in the Josephson junction and microwave devices. Such temperature limits result in dielectric films with high densities of performance-degrading defects.

We suggest that high-purity low-temperature deposited germanium is a viable candidate for dielectric and device passivation layers in multi-layer superconductor circuits. Germanium can be obtained in an even higher purity[11] form than Si, with impurity concentrations as low as $10^8$ $cm^{-3}$. High-resistivity germanium films can be deposited via evaporation at low substrate temperatures, including down to room temperature. If no additional contamination is added during the deposition process, only a few tens of impurity defects would be present in the entire active region of coplanar, microstrip, or stripline structures. It could also potentially replace commonly-used deposited dielectrics with high-loss in microwave and Josephson junction circuit fabrication.

In this study, we measure the microwave and two-level system losses in Nb-based coplanar microwave resonators synthesized on molecular-beam evaporated 1 μm-thick Ge thin-films and on chemical vapor deposited (CVD) 9 μm-thick homoepitaxial Si thin-films. To enhance our sensitivity to the interface and properties of the deposited dielectrics, we have intentionally designed the resonators for this study with narrow trace and gap widths, and minimal (i.e., <20 nm) over-etch. This is in contrast to other studies that focus on maximizing the quality factors by designing resonators with large trace widths (>10 μm) and significant over-etch[12,13].

We deposit amorphous germanium (*a*-Ge) films via thermal evaporation onto undoped float-zone Si (100) substrates with resistivities greater than 10 kOhm-cm. Undoped crystalline Ge wafers were broken into small pieces for use as evaporation material. The Si substrate surface is cleaned in ultrasonic baths of USP-grade acetone, then ethanol for ten minutes, followed by etching for 5 minutes in aqueous 2% HF solution. For Ge deposition, the chamber is evacuated to a base pressure of less than $5\times10^{-9}$ Torr before slowly warming the Ge evaporation source (SVT High Temperature Effusion Cell) to 1400 °C. The Ge is deposited at ≈0.25 nm/sec to a total film thickness of ≈1 μm. Epitaxial Si films are prepared in a commercial foundry at Lawrence Semiconductor Research Laboratory (Tempe, AZ). The substrate is a (100) orientation CZ-grown Si wafer (resistivity ρ = 1000-25,000 Ω-cm), cleaned using an in-situ high purity HCl etch. The ≈ 9 μm thick epi-layer was deposited at 900 °C, using 2% silane in $H_2$, with a deposition rate of ≈ 300 nm/minute. Prior to insertion into the Nb metal layer deposition system, the epi Si/Si film surface is cleaned in ultrasonic baths for 10 minutes—first in semiconductor-grade acetone, then ethanol. To remove residual surface contamination, the film is flash-heated in vacuum[14] ($<5\times10^{-9}$ Torr) to 850 °C, and then held at 650 °C for 1 hour before cooling for about 1 hour to room temperature before Nb sputter deposition.

Nb metal films are sputter-deposited onto the substrates or deposited layers without breaking vacuum (in the case of *a*-Ge) to minimize surface contamination. Sputter deposition is performed at room temperature and 4 mTorr of Ar in a UHV system with an unbaked base pressure $<5\times10^{-9}$ Torr using a 2" diameter magnetron sputter source with 99.95%-pure Nb targets. The sputter power is 225 W with a source-film distance of 15 cm, resulting in a sputter rate of ≈0.6 nm/sec.

Chemical depth profiles are obtained using Time of Flight Secondary Ion Mass Spectrometry (TOF-SIMS). The analysis beam uses $Ga^+$ ions with a 1 kV Cs sputter ion beam for depth profiling. Because SIMS sensitivity factors are only available for the bulk, and differ greatly from these values for the disordered atomic structures near these Nb/semiconductor junctions, all near-interfacial data presented here will be in raw counts normalized to the relative count rate of the host matrix (Nb and Ge) and are not accurate estimates of atomic concentration.

Coplanar-waveguide (CPW) resonators are prepared using standard photolithography and reactive ion etching (RIE) in a $CF_4$ plasma. We pattern devices into a 50Ω (on silicon) coplanar waveguide resonator configuration with 2-16 μm trace width and 2-8 μm gap width, and less than 20 nm over-etch to maximize electric field interaction with the interface and dielectric films under study. The quarter-wave CPW resonators are capacitively coupled to the microwave feedline. Transmission measurements (S21) are made at ≈40 mK in a closed cycle dilution refrigerator as a function of applied power. The input signal to the feedline is attenuated by 40 dB at room temperature and 20 dB, 10dB, and 20dB on the 4 K, still, and mixing chamber plates, respectively. The output signal is buffered by two isolators in series on the mixing chamber plate and another on the 4 K plate. The signal is amplified by a HEMT amplifier at 4 K and a low-noise amplifier at room temperature. The resonance is fit using the diameter correction method to extract $Q_i$ the internal quality factor of the resonator and thus the loss tangent[15] ($\tan\delta_i = Q_i^{-1}$).

Carrier concentrations determined using room-temperature Hall Effect measurements find the deposited Ge film is n-type with $6\times10^{12}$ cm$^{-3}$ net carriers, essentially identical to the original wafer. Electron Paramagnetic Resonance (EPR) measurements on these *a*-Ge films (performed using the in-situ parallel plate EPR technique described in our earlier work[16]) find the paramagnetic defect concentration is below the detection limit of ≈$10^{17}$ cm$^{-3}$. X-ray diffraction characterization, performed using a PANalytical X'pert MRD Pro, of the *a*-Ge films do not exhibit any sharp Bragg diffraction peaks, characteristic of amorphous material. Raman spectroscopy, measured with a 532 nm laser at 0.75 mW with 0.5 μm spot size, show a broad peak near 290 cm$^{-1}$, similar to that found for electrolytically deposited[17] *a*-Ge, as would be expected for an amorphous Ge film.

Si films deposited under the same conditions as used in our study have been measured to be slightly n-type, with net carrier concentrations less than $10^{12}$ cm$^{-3}$ (at room temperature) as measured by spreading probe resistance. SIMS measurements find that the impurity concentration is below the detection limit of $\approx 10^{12}$ cm$^{-3}$.

Table 1 and Figure 1 summarize results from low power, low-temperature microwave measurements on Nb-based CPW resonators on float-zone silicon, homoepitaxial Si films, a high-resistivity Ge wafer, and room-temperature deposited *a*-Ge. The resonator with room-temperature deposited *a*-Ge dielectric exhibits a total dielectric loss comparable to Nb/epi-Si wafer, and the lowest single-photon loss in a deposited amorphous dielectric reported to date. Low two-level system loss densities have also been reported in acoustic and thermal measurements on deposited amorphous dielectrics[18–21]. While acoustic measurements are at a much lower frequency and coupled to strain instead of electric fields, TLS spectroscopy measurements in strain and field find similar spectra and associate these with bulk and surface TLS, respectively[22,23] Our *a*-Ge films deposited at room temperature without any post-processing exhibit single-photon loss tangents 6 times lower than reported values of internal friction on e-beam and sputtered *a*-Ge films by Liu[19], and 2 times lower than *a*-Ge films reported with post-process annealing (5 hours at 350 °C). In those studies, Liu, et al. report that low energy excitations of *a*-Ge are highly dependent on the preparation method, which they attribute to structural differences in the films. Further enhancement may be possible based on later studies by Liu[20], where they report *a*-Si films that do not exhibit two-level system loss even without hydrogen passivation by using the optimized 400 °C substrate growth temperature, achieving Q = 5×10$^5$ (i.e., tan δ = 2x10$^{-6}$).

Comparing our results to those in the surveys of superconducting coplanar resonators[1,2,4], we find that our high-purity deposited *a*-Ge films exhibit loss much lower than has been reported for *a*-Si:H, sputtered *a*-Si, lumped element *a*-SiN$_x$ resonators[7], and other deposited dielectric materials[24]. The single-photon loss in the *a*-Ge film is similar in magnitude to early reports of high-quality resonators on single-crystal Si or sapphire, but higher than more recent reports, including the CPW resonators on float-zone Si reported here. The loss in a superconducting resonator is sensitive to the cleanliness of the interfaces, and the devices presented here are no exception. Following the technique from Ref. [8], the filling factors of the dielectric film and substrate were calculated using the electric field distributions simulated in Ansys Maxwell for each geometry and material. The simulation assumes a 2 nm oxide interface layer is present at every surface and interface with dielectric constants 3.6, 2.5, and 10 for the silicon, germanium, and metal oxides, respectively. Bulk room temperature dielectric constants were used for the silicon 11.9 and germanium 16.0. The effective loss tangents in Table 1were calculated by subtracting the interface loss and substrate loss (for the films only) from the measured single photon loss tangents.

$$\tan \delta_{eff} = \tan \delta_{\langle n \rangle = 1} - \sum FF_i \times \tan \delta_i$$

The interface and bulk silicon loss tangents are taken from Ref. [25]. The effective loss tangent is dominated by the uncertainty in the interface and substrate loss for the films but the calculated value is useful for comparing resonators of different geometries within a material system. Differences between the effective loss tangents within a material system are likely due to differences in interface contamination between the devices. More reliable determination of the material and interface losses requires measuring a large number of resonators with varying geometries[25].

In order to understand the unexpected higher loss in c-Ge than a-Ge, time of flight SIMS depth profiles were collected for the Nb/a-Ge and the Nb/crystalline Ge wafer interfaces. The profiles identify H, C, O, F, and Cl at the interfaces with very low contamination in the a-Ge film or substrate (Figure 2). The TOF-SIMS data are shown with relative yield instead of concentration to avoid problems with matrix effects near the metal/semiconductor interface. The a-Ge (which had the Nb deposited in-situ) has fewer contaminants near the interface when compared to the Nb deposited directly on the air-exposed, chemically cleaned bulk Ge wafer. On the bulk Ge wafer, we observe contamination C, F, and Cl peaks extending about 20, 20, and 40 nm into the Ge, respectively. Based on these results, the lower loss in the deposited a-Ge resonator is not unexpected, given the high chemical purity of the deposited film's interface.

The smaller difference in loss tangent between the epitaxial silicon film and the c-Si substrate is likely due to sample-to-sample variation in the interface preparation. However, the loss tangent of the CZ-Si wafer was not measured. The effective loss tangent calculation assumes $\tan \delta = 2.6 \times 10^{-7}$ but the value of the bulk loss tangent of crystalline silicon at milli-Kelvin temperatures is still an open question as interface losses dominate the measurements[25]. Similarly, the differences in loss between high resistivity CZ and FZ silicon has not been studied.

The 1 μm thick a-Ge layer evaporated onto Si (100) has an RMS roughness of 1.02 Å as measured by atomic force microscopy. Such low roughness topographies are necessary for strict device dimensioning and processing in multi-layer structures. These a-Ge films have another advantage for such applications. They have sufficient conductivity at room temperature to protect electronic devices, such as transistors and Josephson Junctions, from electrostatic discharge during fabrication and storage. Room-temperature deposited Ge films exhibit low loss, as shown here, and do not require elevated temperature thermal processing, enabling multi-layer resonator geometries for use in microstrip and stripline microwave devices, as well as interwiring, isolation, and passivation layers. Although the magnetic properties of Ge have many advantages for all the applications discussed here, the natural form that was used in this study does contain a 7.8% abundance of $^{73}$Ge, with 9/2 nuclear spin, which could potentially contribute to loss and noise in spin sensitive devices[26,27]. Isotopically enriched $^{74}$Ge is available and could be used to make thin films if the nuclear spin loss would contribute significantly to loss[28].

In summary, we have demonstrated that deposited dielectrics in CPW resonators exhibit low loss when operated at single-photon powers and low (≈40 mK) temperatures. CPW resonators on amorphous germanium films deposited at room-temperature have single-photon loss tangents of $1$–$2\times10^{-6}$, and CPW resonators on 9 μm thick high-purity CVD homoepitaxial silicon films have single-photon loss tangents of less than $0.6$–$2\times10^{-5}$. Interface losses presumably dominate the resonator performance in these devices.

These results show that room-temperature deposited amorphous Ge layers and CVD Si films exhibit microwave properties suitable for incorporation in quantum computing and sensing applications. The room-temperature deposited amorphous Ge layers could also be used to make co-planar, microstrip and stripline resonators, or could be utilized in place of the currently-used $SiO_x$ and $Si_xN_y$ for dielectric isolation, wiring, and passivation layers.

Data are available upon reasonable request to the authors and with the permission of Northrop Grumman Corporation.

| Dielectric Material | Sample | Width μm | Gap μm | $f_0$ GHz | tan δ ⟨n⟩ = 1 | Filling Factors | | | | | tan $\delta_{eff}$ |
|---|---|---|---|---|---|---|---|---|---|---|---|
| | | | | | | Film | Sub | MS ×10$^{-2}$ | SA ×10$^{-4}$ | MA ×10$^{-4}$ | |
| a-Ge film (1 μm) | 1 | 2 | 2 | 6.31 | 1.1×10$^{-5}$ | 0.63 | 0.27 | 1.2 | 6.3 | 2.5 | 4.9×10$^{-6}$ |
| | 2 | 16 | 8 | 7.30 | 1.3×10$^{-5}$ | 0.24 | 0.68 | 0.29 | 1.4 | 0.48 | 4.7×10$^{-6}$ |
| Bulk c-Ge (Substrate) | 3 | 2 | 2 | 5.38 | 2.7×10$^{-4}$ | – | 0.91 | 1.3 | 6.1 | 2.4 | 2.9×10$^{-4}$ |
| | 4 | 4 | 2 | 5.81 | 7.5×10$^{-5}$ | – | 0.92 | 1.0 | 4.7 | 1.8 | 7.5×10$^{-5}$ |
| | 5 | 8 | 4 | 6.33 | 5.9×10$^{-5}$ | – | 0.93 | 0.56 | 2.5 | 0.92 | 6.0×10$^{-5}$ |
| CVD Epi-Si film (9 μm) | 6 | 2 | 2 | 4.70 | 1.8×10$^{-5}$ | 0.89 | 0.0037 | 0.69 | 8.4 | 2.6 | 1.4×10$^{-5}$ |
| | 7 | 2 | 2 | 6.46 | 1.3×10$^{-5}$ | 0.89 | 0.0037 | 0.69 | 8.4 | 2.6 | 8.0×10$^{-6}$ |
| | 8 | 4 | 2 | 7.28 | 1.2×10$^{-5}$ | 0.89 | 0.012 | 0.54 | 6.5 | 1.9 | 9.1×10$^{-6}$ |
| | 9 | 8 | 4 | 5.87 | 6.0×10$^{-6}$ | 0.85 | 0.056 | 0.30 | 3.5 | 0.96 | 4.3×10$^{-6}$ |
| | 10 | 8 | 4 | 8.33 | 8.2×10$^{-6}$ | 0.85 | 0.056 | 0.30 | 3.5 | 0.96 | 6.83×10$^{-6}$ |
| c-Si substrate | 11 | 5 | 4 | 6.79 | 3.5×10$^{-6}$ | – | 0.91 | 0.35 | 4.1 | 1.1 | 8.34×10$^{-7}$ |

Table 1: Summary of results for coplanar microwave resonators measured at single-photon power. The filling factors are for the film, substrate, and the metal-substrate (MS), substrate-air (SA), and metal-air (MA) interfaces. The single-photon loss tangent is taken from the measured loss tangent at the single-photon occupation ⟨n⟩ = 1 in the resonator. The effective loss tangent tan $\delta_{eff}$ is calculated by subtracting the interface losses from the single photon loss as described in the text. The interface loss tangents for the MS (4.8×10$^{-4}$), SA (1.7×10$^{-3}$), and MA (3.3×10$^{-3}$) are taken from Ref [25.]. For the films, tan $\delta_{eff}$ is calculated by subtracting the silicon substrate loss (tan δ = 2.6 × 10$^{-7}$). For the c-Ge and c-Si substrates, tan $\delta_{eff}$ is calculated for the substrate.

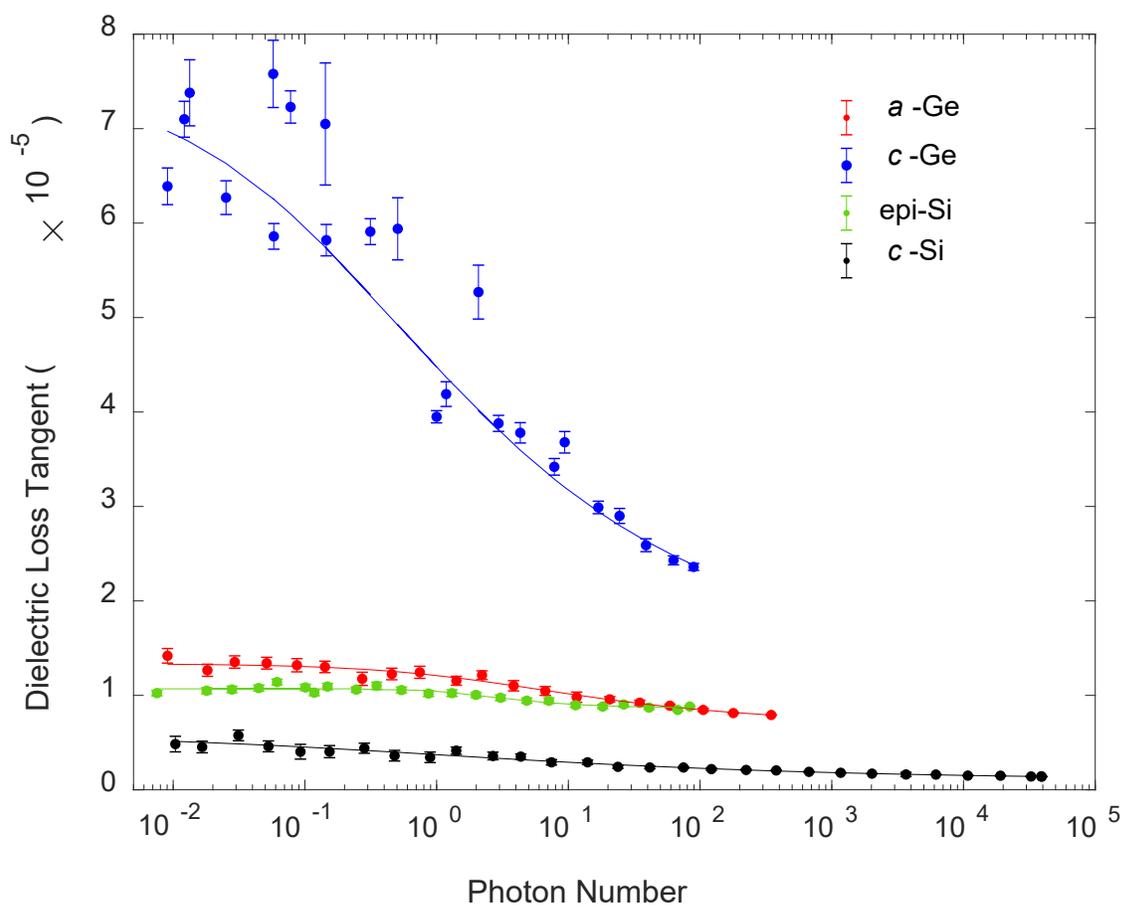

Figure 1: Dielectric loss tangents for niobium CPW devices (samples 1, 4, 8, 11) measured at ≈40 mK shown as a function of photon number. The selected resonators have the lowest effective loss of each material system. The lines are fits to the two-level system model[29,30].

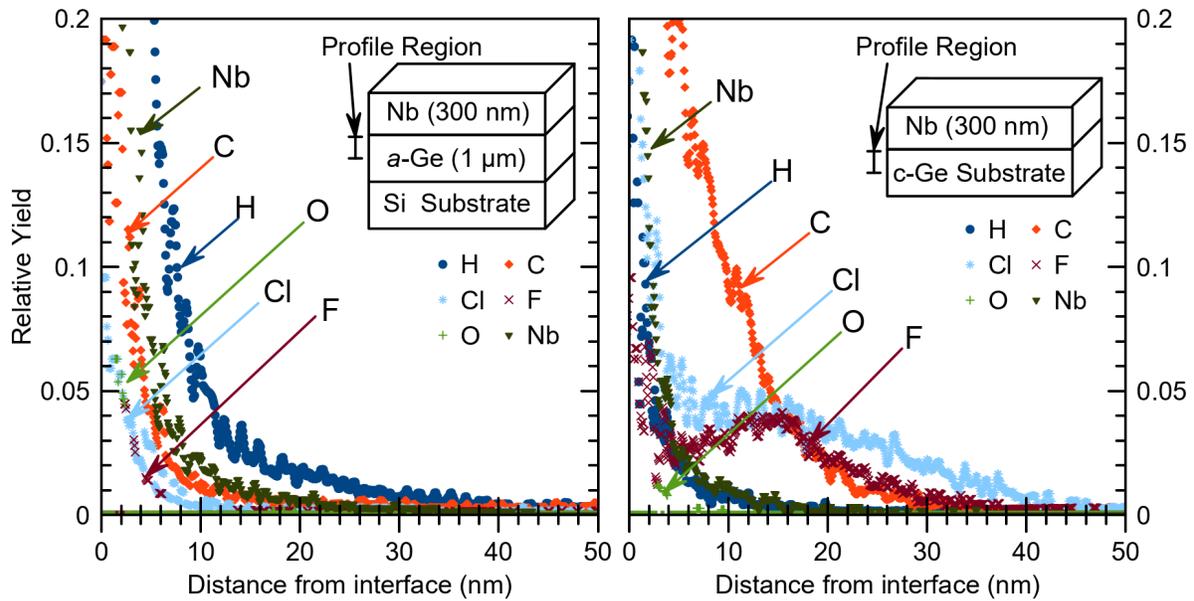

Figure 2: TOF-SIMS depth profiles showing uncorrected relative intensity for elements detected near the Nb/Ge interface (H, C, O, F, and Cl) beginning at 300 nm from the top surface (the depth of the Nb/Ge interface). The room-temperature deposited amorphous Ge (left) has very little interface contamination, with H the only other significant detection past 20 nm. The specimen deposited on single-crystal Ge wafer (right) has a significant interface C peak, F extending about 30 nm deep, with Cl extending about 45-50 nm into the Ge layer.